# Sub-wavelength localized all-optical helicity-independent magnetic switching using plasmonic gold nanostructures

Themistoklis Sidiropoulos[1,†], Puloma Singh[1,†], Tino Noll[1], Michael Schneider[1], Dieter Engel[1], Denny Sommer[1], Felix Steinbach[1], Ingo Will[1], Bastian Pfau[1], Clemens von Korff Schmising[1], Stefan Eisebitt[1]

[1]Max Born Institute for Nonlinear Optics and Short Pulse Spectroscopy, 12489 Berlin, Germany

**Abstract**

All-optical helicity-independent switching (AO-HIS) is of interest for ultrafast and energy efficient magnetic switching in future magnetic data storage approaches. Yet, to achieve high bit density magnetic recording it is necessary to reduce the size of the magnetic bits addressed by laser pulses at well-controlled positions. Metallic nanostructures that support localized surface plasmons enable spatial electromagnetic confinement well below the diffraction limit and rare-earth transition metal alloys such as GdTbCo have demonstrated nanometre-sized stable domains. Here, we deposit plasmonic gold nanostructures on a GdTbCo film and probe the magnetic state using magnetic force microscopy. We observe localized AO-HIS down to a critical dimension of 240 nm after excitation of the gold nanostructures by a single 370 fs long laser pulse with a centre wavelength of 1030 nm. We demonstrate that the strong localization of optical fields through plasmonic nanostructures enables reproducible localized nanoscale AO-HIS at sub-wavelength length scales. We study the influence of the localized electromagnetic field enhancement by the plasmonic nanostructures on the required fluence to switch the magnetization.

The observation of ultrafast magnetization reversal in a ferrimagnetic alloy after excitation with a single intense ultrashort optical pulse has sparked the vision of novel ultrafast and efficient magnetic storage devices [1,2]. The all-optical helicity-independent switching (AO-HIS) of the magnetization state in ferrimagnetic alloys is understood by different demagnetization dynamics of the two antiferromagnetically coupled spin sub-lattices [3]. Excitation with an ultrashort laser pulse leads to an increase in the electron temperature and triggers the demagnetization of the sub-lattices on their own time-scales as exchange coupling between both lattices is suppressed. Eventually exchange coupling is restored and conservation of the angular momentum leads to a reversal of the magnetic state within picoseconds. While the ultrashort timescale of AO-HIS has been demonstrated, the reduction of the switched areas towards dimensions typically used in commercial magnetic storage devices still remains challenging.

† These authors contributed equally

Fundamental limitations in the size of the switched areas arise from the diffraction limit for focussing the excitation laser, the minimal stable domain size depending on the magnetic medium, but also from the specifics of heat transport on the nanometre scale due to electron diffusion [4]. Ferrimagnetic alloys containing rare earth elements with high magnetic anisotropy such as Tb have shown to lead to stable domain sizes on the order of 100 nm, significantly smaller compared to alloys with only one rare earth element with low magnetic anisotropy [5,6]. AO-HIS of areas that are below the diffraction limit for an excitation in the visible regime can be achieved through structuring the magnetic film or through transient gratings at extreme ultraviolet photon energies [4,7–9]. An alternative approach is to deposit metallic nanostructures that support localised plasmons on the magnetic film, as illustrated in Figure 1(a). Localised surface plasmons are capable of confining light to sub-wavelength scales and thus can enable the switching of sub-wavelength magnetic areas under the right conditions.

Two-dimensional numerical simulations along the excitation direction with the finite-element solver COMSOL shown in Figure 1(b), highlight the extreme localisation of the electric field near the gold nanostructure after optical excitation with 1030 nm light. Nanoparticles supporting localised surface plasmons can efficiently absorb and scatter the incoming radiation that consequently leads to a large dissipated power in the magnetic layer [10,11]. Thus, plasmonic nanostructures have the potential for a reduction in the critical dimensions of switched magnetic areas while also enabling switching at a reduced laser power. Indeed, reduction in the required fluence to observe switching in plasmonic nanoparticle arrays has been demonstrated [12–14]. The controlled (sub-wavelength) localisation of AO-HIS using metallic nanostructures has been demonstrated in x-ray imaging experiments, relying on phase reconstruction techniques currently only possible at large scale facilities [5,15].

Here, we exploit the high spatial resolution of magnetic force microscopy (MFM) in lab-based environments [16,17] to probe the sub-wavelength localisation of AO-HIS through the excitation of plasmonic nanoparticles and to study the effect of light scattering and absorption for particular nanoparticle arrangements on the switched magnetization sate.

[Type here]

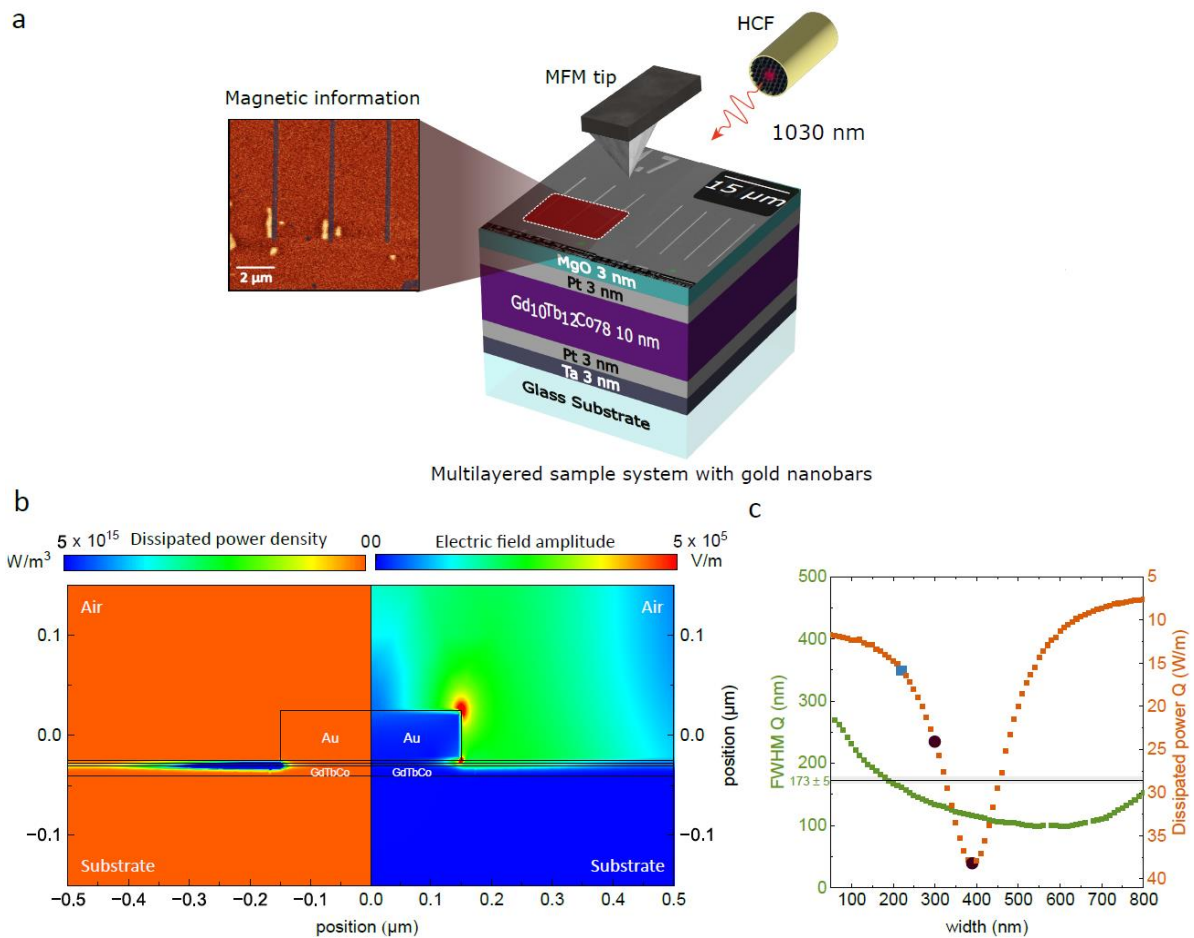

**Figure 1** Schematic of plasmonic magnetic multilayered sample system along with 2D COMSOL simulations. (a) Schematic of sample composition and the setup to study AO-HIS. Gold nanobars are deposited on a Ta(3nm)/Pt(3nm)/Gd$_{10}$Tb$_{12}$Co$_{78}$(10nm)/Pt(3nm) magnetic alloy multilayer separated by a 3nm thick MgO layer. A hollow-core fibre (HCF) delivers ultrashort laser pulses at 1030nm that trigger AO-HIS. The change in the magnetization state is probed from the top by MFM (b) 2D simulation along the excitation direction of the maximum electric field amplitude (right panel) highlighting the strong localization to the edges of the nanostructure. The dissipated power (left panel) in the magnetic layer highlights the efficient scattering of the nanoparticle. (c) The integrated dissipated power (right axis) in the magnetic layer becomes the largest for a nanoparticle of 390nm width. Red dots and the blue square mark the dimensions of the studied nanodiscs and nanobars, respectively. The left axis shows the width of the dissipated power profile in the magnetic layer as determined from a Gaussian fit. The horizontal line indicates the minimal stable domain size.

We have fabricated a 10 nm thick ferrimagnetic Gd$_{10}$Tb$_{12}$Co$_{78}$ alloy by DC magnetron sputtering, sandwiched between two 3 nm thick layers of Pt. The high concentration of Tb facilitates the stability of smaller domain sizes due to an increase in magnetic anisotropy which is reflected in the large out-of-plane magnetization with a coercive field of ~450 mT. From

[Type here]

laser-induced domain patterns in the sample, we identify the smallest isolated stable magnetic domain size in the film to be 173 ± 5 nm (not shown). An additional 3 nm MgO cap layer acts as a spacer layer between the Pt cap and the Au nanostructures and prevents quenching of the plasmon resonance. Plasmonic gold nanobars with a length of 20 µm and a width of 220 nm, as well as gold nanodiscs with a diameter of 300 nm and 400 nm, have been fabricated through electron beam lithography and a subsequent deposition of 30 nm gold on the magnetic sample system. The nanostructures are separated by 3 µm in order to avoid direct coupling.

In Figure 1(c) we show the calculated total dissipated power in the magnetic layer that becomes the largest for a 390 nm wide nanoparticle, with a FWHM of the power distribution of about 150 nm. The width of the nanostructures was chosen to be in the vicinity of this resonance condition. While the off-resonant excitation of the nanobars reduces optical confinement, it can nevertheless be particularly efficient when it relates favourably to the minimal stable domain sizes. Indeed, from a Gaussian fit to the calculated profile of the dissipated power density in the magnetic layer for different nanobar widths, we find that for a 220 nm wide bar, the dissipated power distribution is similar to the minimal stable domain size. The precise localized surface plasmon resonance condition, however, has a strong dependence on the exact shape of the nanostructure and the excitation condition [18].

The magnetic state of the sample is probed with an enclosed MFM system (Bruker Dimension Icon) that does not allow for a free-space coupling of the laser beam onto the sample for in-situ operation. Therefore, we developed a fibre-based optical setup that delivers sub-1-ps laser pulses at 1030 nm with few mJ/cm$^2$ energy density sufficient to observe AO-HIS, enabling studies with a fast turnaround time. To transport light into the encapsulated system, we use a hollow-core photonic crystal fibre (GLO PMC-C-Yb-7C), where light is guided in a 60 µm diameter inner air core. The use of a hollow-core fibre eliminates nonlinearities from an otherwise solid core and provides an intense ultrashort laser pulse at the output of the fibre [19]. To this end, we focus the output of the laser (Amplitude Satsuma HP) onto the fibre's end-facet to match the fibre's mode-field diameter of around 45 µm. We achieve more than 81% transmission of 370 fs long pulses with up to 800 nJ without significant pulse distortions and without damaging the fibre. The out-coupled laser pulses are then focused onto the sample near the MFM tip. The laser spot diameter on the sample is 62 µm as characterized by the second harmonic emission from a Boron Nitrate film. The delivery of these ultrashort pulses through the hollow-core fibre into the encapsulated MFM enables us to irradiate the magnetic sample from the top with an ultrashort intense laser pulses and directly image the magnetic state in-

[Type here]

situ. The spatial resolution of the images recorded is controlled by the scan step size and limited by the MFM tip. Here, we perform scans with 256 pixels per line and use a hard-magnetic coated silicon tip (Nanosensors SSS-MFMR) that has a nominal resolution of 25 nm.

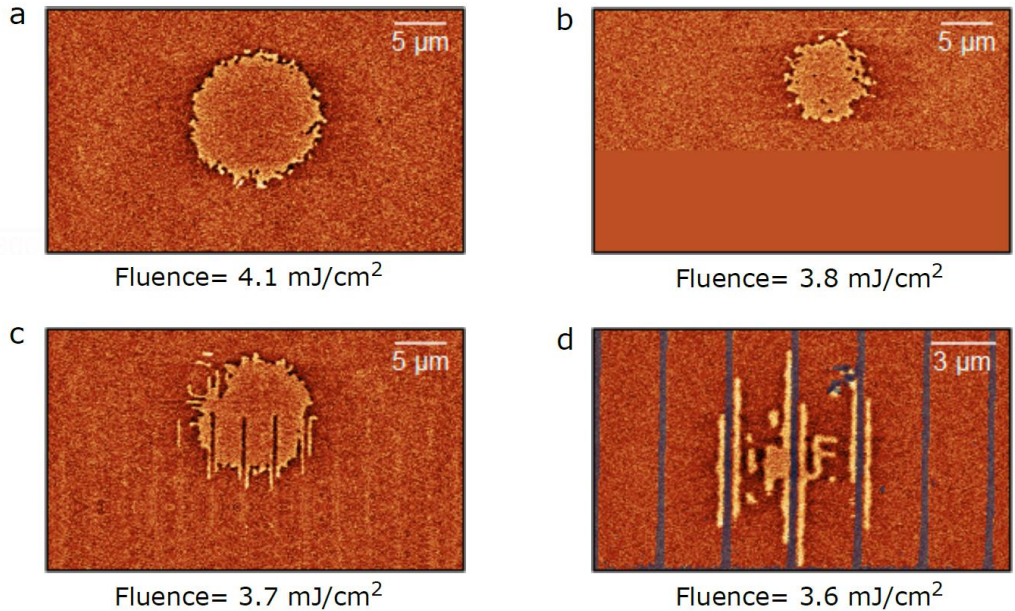

**Figure 2** Determination of the threshold fluence required for localised AO-HIS. A single 370 fs short 1030 nm pulse with a fluence of (a) 4.1 mJ/cm$^2$ and (b) 3.8 mJ/cm$^2$ lead to AO-HIS on the bare sample and allow determining the threshold fluence for switching. (c) Below threshold excitation of nanobars with 3.7 mJ/cm$^2$. (d) Reduction of fluence to 3.6 mJ/cm$^2$ results in localised switching with far-field scattering leading to switched areas between the nanobars (blue lines).

We first verify that the delivered pulses can trigger a switching of the magnetization state, and therefore irradiate the bare sample with a laser fluence of 4.1 mJ/cm$^2$. Magnetic force microscopy is sensitive to the force gradient acting on the magnetic tip and in Figure 2(a) we show the recorded magnetization state as the phase shift in the cantilever's resonance [20,21]. Immediately apparent is the central region of the laser spot where the fluence is sufficient to switch the magnetization. We then gradually decrease the fluence, thereby reducing the area of magnetic switching, until no reversal of magnetization is observed on the bare sample. Figure 2(b) shows the switched state after excitation with 3.8 mJ/cm$^2$; a further reduction in fluence no longer causes switching. This threshold fluence of 3.7 mJ/cm$^2$ is consistent with previous studies on the magnetic multilayer alloy with the same composition [6].

Next, we move to a sample area with plasmonic gold nanobars and irradiate the film with a laser polarization perpendicular to long axis of the bars and the determined threshold fluence of 3.7 mJ/cm$^2$. The switched area shown in Figure 2(c) is comparable to that observed

[Type here]

at the bare sample when the irradiated fluence was 4.1 mJ/cm$^2$. This is a clear indication that the plasmonic nanostructures reduce the minimal required fluence to reverse the magnetization, especially when considering that AO-HIS is strongly non-linear in its fluence dependence and hence a very sudden low-fluence threshold behaviour is typical[22]. Interestingly, the edges of the switched spot show distinct line-like features adjacent to the nanobars, indicating that it is the plasmonic confinement that reduces the required fluence to switch the magnetization state. In line with that observation, a further reduction in the fluence to 3.6 mJ/cm$^2$ enables switching of the magnetization state in an area that is well confined to the nanobar, as shown in Figure 2(d). Surprisingly, the switched areas do to some extent extend between the nanobars. We attribute this behaviour to far-field scattering under the off-resonant excitation condition of the localized plasmons.

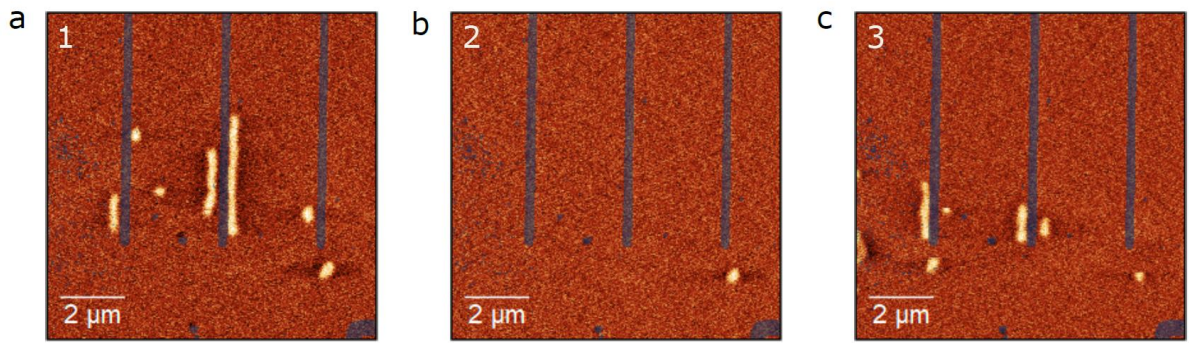

**Figure 3** Spatially localized all optical switching obtained from a single 370 fs short 1030 nm laser pulse with 3.5 mJ/cm$^2$ incident fluence irradiated on gold nanobars highlighted by blue lines. (a) First shot focussed on the plasmonic nanobars leads to a switching of the magnetization in the vicinity of the nanobars. (b) The second laser shot switches the magnetization state back to its original state. (c) After excitation with the third laser pulse in the same location, the magnetization state is reversed again at the edge of the nanobars.

Therefore, we reduce the fluence further to 3.5 mJ/cm$^2$ to push also these areas below the threshold fluence by reducing scattering and to highlight the strong spatial confinement. As seen in Figure 3, we now observe switching exclusively at the edges of the nanobars where plasmonic confinement is largest, in line with the simulations presented in Fig. 1(b). In Figure 3, we demonstrate AO-HIS with single ultrashort laser pulses with 3.5 mJ/cm$^2$ incident fluence, where each single pulse consecutively toggles the state of magnetization. After excitation with the first laser pulse, switching of the magnetic state occurs in the vicinity of the nanobars, Figure 3(a). The state of magnetization is reversed after the next consecutive shot when the laser excites the same position, Figure 3(b). After the next single pulse irradiation, the magnetic switching reappears at the edges of the nanobars, Figure 3(c). While switching does not occur at precisely the same position in our geometry with long nanobars, there is a strong overlap in the switched areas. We attribute the deviation from perfect toggle reversal of the magnetic domains to small shifts in the laser spot position and to small fluctuations in the laser intensity. As the fluence is very close to the switching threshold, small intensity changes in the nonlinear process can have significant impact on the switched areas and lead to the slight differences between the two shots shown in Figure 3(a) and Figure 3(c).

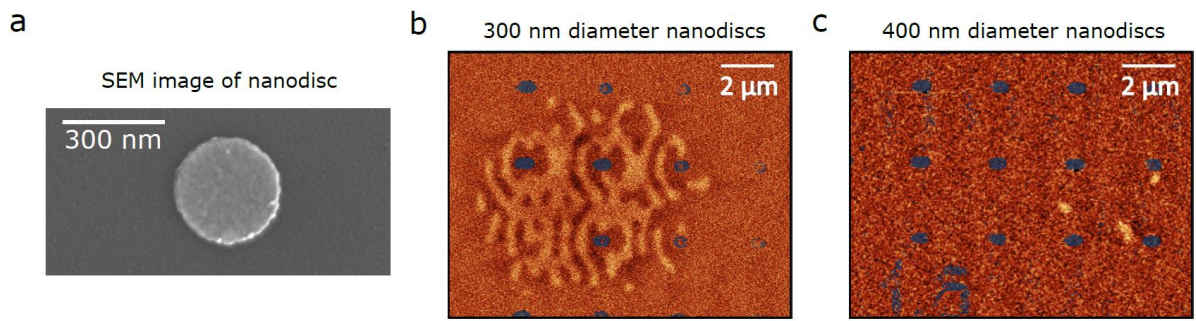

**Figure 4** AO-HIS from off- and on-resonant nanodiscs. (a) Scanning electron microscope (SEM) image of a gold nanodisc with a diameter of 319 nm. (b) Magnetization state observed after laser irradiation with an incident fluence of 3.7 mJ/cm$^2$ on the 300 nm diameter and (c) 400 nm diameter gold nanodiscs, whose topography are depicted by blue dots

The off-resonant nanobars already achieve a switched area of 240 nm width and 1-2 µm length. The use of shorter nanobars or smaller structures such as gold nanodiscs as shown in Figure 4(a), has the potential to reduce the size of both critical dimensions of the quasi-2D domains and hence reduce the size of the switched area. As seen in Figure 4 (b), excitation of 300 nm diameter nanodiscs at 3.7 mJ/cm$^2$ leads to a dipole like pattern of the switched magnetization around the individual nanodiscs with a periodic modulation of the magnetization away from the nanodiscs reminiscent of an interference pattern. We attribute the occurrence of such magnetization patterns to the dipole-like radiation of off-resonant excited nanodiscs that interferes with the emission from neighboring particles., The efficient forward scattering for the 300nm nanodiscs that interferes constructively and destructively is absorbed in the magnetic layer and becomes relevant when the resulting local fluence approaches the threshold fluence. Interestingly, the excitation of 400 nm diameter nanodiscs that are close to the optimum plasmonic resonance condition, show no effect of far- field scattering and hardly any switching in the magnetization at the same incident fluence, see Figure 4 (c). We speculate that at on-resonance condition, where strong near-field coupling dominates over far-field scattering, leads to a localization of the dissipated power smaller than the smallest stable domain size in

[Type here]

our magnetic layer. If confirmed, this would open the straightforward way to increased spatial confinement with suitably optimized magnetic media.

In conclusion, we have observed sub-wavelength spatial confinement in single laser pulse all-optical switching down to domains with critical dimensions of about $\lambda/4$. In our magnetic medium, this coincides with the minimum stable domain width, suggesting scalability to critical dimensions of below 100 nm with a suitable magnetic medium. We observe a reduction in the required threshold fluence for such switching via plasmonic resonances and see single shot toggle switching for subsequent laser shots. For plasmonic nanodiscs at suitable laser fluence we observe specific switching patterns which we attribute to interference in the scattered light field of neighbouring nanodiscs. The constructive (destructive) interference raises (decreases) the local fluence above (below) the threshold fluence for switching. Imprinting the scattered light field into the magnetisation state, observed for the first time to our knowledge, opens new pathways to exploiting plasmonic structures in densely packed geometries in order to achieve a high areal density of optically switched regions. Here, well-chosen non-symmetric arrangements may be beneficial, and warrant further theoretical study. Finally, we note that in-situ studies of laser-induced magnetic switching phenomena with a spatial resolution well below that of magneto-optical microscopes has been made possible by coupling fs laser pulses into a commercial magnetic force microscope via a hollow-core fibre, allowing for efficient, laboratory-based experiments in this field.

## Acknowledgments

C.v.K.S., and S.E. would like to thank the German Research Foundation (DFG) for funding through CRC/TRR 227 projects A02 (project ID 328545488).